\documentclass[a4paper,aps,prd,preprint,groupedaddress,nofootinbib]{revtex4-1}
\usepackage{amsmath}
\usepackage{amsfonts}
\usepackage{graphicx}
\usepackage{subfigure}
\usepackage{epsfig}
\usepackage[vcentermath]{youngtab}

\def\bea{\begin{eqnarray}}
\def\eea{\end{eqnarray}}
\def\be{\begin{equation}}
\def\ee{\end{equation}}
\def\Nm{\mathcal{N}}
\def\Lm{\mathcal{L}}
\def\l{\lambda}
\def\ph{\phantom}

\begin{document}
\preprint{YITP-SB-11-22
}
\title{Orientifold daughter of $\mathcal{N}=4$ SYM and double-trace running}
\author{Pedro Liendo\footnote{Email: pedro.liendo@stonybrook.edu}}
\affiliation{C. N. Yang Institute for Theoretical Physics,
\it Stony Brook University, \\
\it Stony Brook, NY 11794-3840, USA}

\date{\today}

\bigskip

\begin{abstract}
We study the orientifold daughter of $\Nm=4$ super Yang-Mills as a candidate non-supersymmetric large~$N$ conformal field theory. In a theory with vanishing single-trace beta functions that contains scalars in the adjoint representation, conformal invariance might still be broken by renormalization of double-trace terms to leading order at large~$N$. In this note we perform a diagrammatic analysis and argue that the orientifold daughter does not suffer from double-trace running. This implies an exact large~$N$ equivalence between this theory and a subsector of $\Nm=4$ SYM.
\end{abstract}

\maketitle

\section{Introduction}

\label{intro}

	Conformal field theories play a prominent role in theoretical physics. In four dimensions, it is easy to 
	find \textit{supersymmetric} CFTs, however, constructing interacting conformal field theories in the absence of SUSY seems 
	to 
	be a harder task.
	An early attempt for the construction of such theories was to use the AdS/CFT 
	correspondence \cite{Maldacena:1997re,Witten:1998qj,Gubser:1998bc} for orbifolds of $\Nm=4$ SYM 
	\cite{Kachru:1998ys,Lawrence:1998ja}. These 
	are constructed by 
	placing a 
	stack of $N$ D3-branes at an orbifold singularity $\mathbb{R}^6/\Gamma$ where $\Gamma$ is a discrete subgroup of the R-symmetry.  
	Inheritance principles \cite{Bershadsky:1998mb,Bershadsky:1998cb}, 
	then, guarantee that the beta functions of marginal single-trace operators vanish in the large $N$ limit. If the orbifold group	
	$\Gamma \not \subset SU(3)$ supersymmetry is completely broken and a potential conformal field 
	theory with reduced supersymmetry is obtained.
		
	Now, whenever there are scalars in the 
	adjoint or bifundamental representation, there is a logarithmic running of double-trace operators present in the 
	quantum effective action \cite{Tseytlin:1999ii,Csaki:1999uy,Adams:2001jb,Dymarsky:2005uh,Dymarsky:2005nc}
	\be 
	\label{deltaS}
	\delta S = -f \int d^4x \mathcal{O} \bar{\mathcal{O}}\,.
	\ee	
	This is a \textit{leading} effect at large $N$. 
	While for supersymmetric orbifolds one can always tune the double-trace couplings
	to their conformal fixed points, for  non-supersymmetric orbifolds
	the double-trace beta functions have complex zeros and conformal invariance is {\it always}
	broken \cite{Dymarsky:2005uh, Dymarsky:2005nc}.	
	
    The authors of \cite{Dymarsky:2005nc} also found a non-trivial one-to-one correspondence between the breaking of 
    conformal invariance in the field theory and the presence of closed string tachyons in the twisted sector of the dual string 
    theory. This result is somewhat
    surprising because the correspondence is between perturbative gauge theory (dual to strongly curved AdS) and
    \textit{flat space} tachyons, indicating that the breaking of conformal invariance may be read from string theory before taking
    the decoupling limit.   
    These results were revisited in \cite{Pomoni:2008de} where it was found that the double-trace beta functions are quadratic in 
    the coupling to all orders in planar perturbation theory.  
        
    Here we will concentrate on the \textit{orientifold} daughter of $\mathcal{N}=4$ SYM
    \cite{Blumenhagen:1999uy,Blumenhagen:1999ns, Angelantonj:1999qg}. This theory arises as the low 
    energy description of D3-branes in a non-tachyonic orientifold of Type 0B, in which the
    tachyon in the original string theory has been projected out by a
    clever choice of the parity operator \cite{Bianchi:1990yu, Sagnotti:1995ga, Sagnotti:1996qj}.
    
    In \cite{Angelantonj:1999qg,Armoni:2007jt} it was 
    argued that this theory is planar equivalent to $\Nm=4$ SYM.    
    In light of the results obtained for orbifolds, the absence of a tachyon in the flat space string theory is a good indication 
    that the orientifold daughter should not suffer from double-trace running and, therefore, be an example of a non-supersymmetric 
    conformal field theory. 
    
    In this note we perform a 
    diagrammatic analysis to see if this theory suffers from double-trace running or not. One possible outcome is that perturbative 
    renormalizability will force us to add double-trace couplings of the from (\ref{deltaS}).
    If the double-trace beta functions have real 
	zeros, conformal invariance can be recovered if we tune the new couplings to their fixed points. This would imply that we have a 
	fixed line passing through the origin of the coupling constant space. On the other hand, if one or more zeros are complex, 
	conformal invariance is broken and the theory is unstable.

    Another possible outcome is that the are no leading double-trace contributions in the effective action. If this is the case, 
    there will be no 
    logarithmic running and it would imply an exact equivalence between the orientifold and a subsector of $\Nm=4$ SYM. 
    Conformal 
    invariance will be preserved but in a rather trivial sense.    
    Our results indicate that this last behavior is the one that 
    characterizes the orientifold daughter. 
    In Section 2 we briefly review how the orientifold theory is constructed. In Section 3 we perform a diagrammatic analysis and 
    show that for each double-trace diagram in $\Nm=4$ there is an analogous diagram in the orientifold 
    daughter and vice versa. Finally, we present our concluding remarks in Section 4.

\section{Orientifold construction}

\label{pre}

    The field theory in which we are interested in is a
    non-supersymmetric $SU(N)$ gauge theory that arises as the low energy
    description of D3-branes in a non-tachyonic orientifold of Type 0B.

    \subsection{Non-tachyonic $\Omega^{\prime}$ projection}

    The Type 0B modular invariant partition function is given by the following string states
    \begin{center}
    (NS$-$,NS$-$)$\oplus$(NS+,NS+)$\oplus$(R$-$,R$-$)$\oplus$(R+,R+)\,.
    \end{center}
    As is well known there is a tachyonic state coming from the
    (NS-,NS-) sector.
    This theory admits more than one consistent orientifold projection
    characterized by different definitions of the parity operator 
    $\Omega$ \cite{Bianchi:1990yu, Sagnotti:1995ga, Sagnotti:1996qj}.
    The parity operator that gives a non-tachyonic orientifold is usually denoted by
    $\Omega^{\prime}=\Omega(-1)^{f_R}$, where $f_R$ is the right world-sheet fermion number. The projected theory 
    $0\textrm{B}/\Omega^{\prime}$ is called $0^{\prime}$B.
    This theory has no tachyon and is similar to the bosonic sector of Type IIB. At the massless level it has a complete set of 
    R-R fields, a graviton and a dilaton.
 
     \subsection{Orientifold daughter}

    If we T-dualize in six directions we obtain the $\Omega(-1)^{f_R}\mathcal{I}_6$
    orientifold of Type 0B \cite{Blumenhagen:1999uy,Blumenhagen:1999ns}, where $\mathcal{I}_6$ is an inversion operator ($x_i 
    \rightarrow -x_i$, $i$=4,...,9). This theory contains an
    orientifold O$^{\prime}3$ plane at $x_4=...=x_9=0$.
    The gauge theory describing $N$ D3-branes in the presence of the
    O$^{\prime}3$ plane is the orientifold daughter of $\mathcal{N}=4$ SYM we
    want to study. Its field content is given by
    
    \begin{center}
    \begin{tabular}{l| c }
                   & $SU(N)$    \\
    \hline
    Vector         &  adj.       \\
    Scalars        &  adj.       \\
    Weyl Fermions  & \tiny{$\yng(1,1)$ + $\overline{\yng(1,1)}$}\\
    \hline
    \end{tabular}
    \end{center}

\begin{figure}[h]
             \begin{center}
            
             \subfigure[\hspace{0.1cm}$\Nm=4$ SYM]{
             
            \includegraphics[scale=0.8]{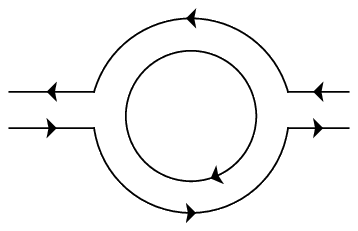}
                       }   
            \hspace{1cm}          
             \subfigure[\hspace{0.1cm}Orientifold]{ 
                      
              \includegraphics[scale=0.8]{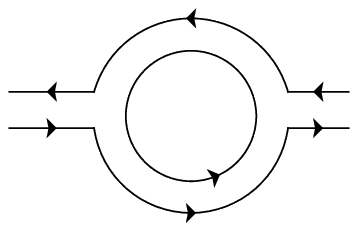}
                     }
              \caption{Fermionic contribution to the scalar self-energy. The only difference is the orientation of the arrows.}
              \label{SelfLoops}
            \end{center}
            \end{figure}

    This theory is very similar to $\mathcal{N}=4$
    SYM, there are 6 real scalars in the adjoint representation and
    4 Dirac fermions in the antisymmetric representation of the gauge group.
    Its planar diagrams are the same as those in
    the parent theory, using a double-line notation \cite{'tHooft:1973jz} is clear that
    the only difference is in the orientation of the color
    arrows for diagrams involving fermions (see Figure \ref{SelfLoops}). This suggests that the orientifold daughter and the
    parent theory are equivalent in the large $N$ limit.
    However, we have to be careful with potential double-trace
    terms  as they might render the
    theory non-conformal.

    We know that in geometric orbifolds of $\mathcal{N}=4$ SYM
    there is a one-to-one correspondence between the presence of
    tachyons in the flat space string theory
    and the breaking of conformal invariance \cite{Dymarsky:2005nc}.
    The absence of a tachyon in the flat space construction of the orientifold field theory is then
    encouraging. Still, we feel an explicit analysis is necessary in order to check whether this theory
    suffers from double-trace running or not.

\section{Diagrammatic analysis}

	Here we consider consider double-trace contributions to 
	the effective action for the orientifold daughter. We will show that they cancel by comparing them with 
	the 
	respective diagrams in $\Nm=4$, which we know does not suffer from double-trace running. 
	\subsection{One-loop diagrams}
	The lagrangian of $\Nm=4$ SYM is well known,
   
        \begin{eqnarray}
        \label{LagrangianN4}
        \mathcal{L} & = & N\bigg( -\frac{1}{2}F_{\mu \nu}F^{\mu
        \nu}+i\bar{\psi}^a \bar{\sigma}^{\mu}D_{\mu}\psi_a
        -D_\mu X^I D^\mu X^I
        \nonumber\\
        & & {}
        +2\sqrt{\lambda}C^{Iab}\psi_{a}X^I\psi_b
        +2\sqrt{\lambda}\bar{C}^{I}_{ab}\bar{\psi}^{a}X^I\bar{\psi}^b+
        \frac{\lambda}{2}[X^I,X^J]^2\bigg)\,.
        \end{eqnarray}
        
        The orientifold daughter is
        obtained through the replacement
        \begin{equation}
        \psi_{i}^{\phantom{k}j} \rightarrow \{\xi_{[ij]},
        \eta^{[ij]}\}\,,
        \end{equation}
        where $i,j$ are color indices \cite{Armoni:2004ub}. Its lagrangian is
        \begin{eqnarray}
        \label{LagrangianOrienti}
        \mathcal{L} & = & N\bigg(-\frac{1}{2}F_{\mu \nu}F^{\mu
        \nu}+\frac{i}{2}\bar{\eta}^a \bar{\sigma}^{\mu}D_{\mu}\eta_a+\frac{i}{2}\bar{\xi}^a \bar{\sigma}^{\mu}D_{\mu}\xi_a
        -D_\mu X^I D^\mu X^I
        \nonumber\\
        & & {}
        +2\sqrt{\lambda}C^{Iab}\xi_a X^I \eta_b + 2\sqrt{\lambda}\bar{C}^{I}_{ab}\bar{\eta}^a X^I \bar{\xi}^b+
        \frac{\lambda}{2}[X^I,X^J]^2 \bigg)\,.
        \end{eqnarray}      
        The color structure of the fermion propagators is as 
        follows,
        \bea 
        \langle \psi_i^{\ph{i}j} \bar{\psi}_k^{\ph{k}l} \rangle & \sim & \delta^l_i \delta^j_k\,,
        \\
        \langle \xi_{[ij]} \bar{\xi}^{[kl]} \rangle & \sim & \delta^l_i \delta_j^k-\delta^k_i \delta^l_j\,,
        \\
        \langle \eta^{[ij]} \bar{\eta}_{[kl]} \rangle & \sim & \delta_l^i \delta^j_k-\delta_k^i \delta_l^j\,.
        \eea
        
          \begin{figure}[h]
             \begin{center}
            
             \subfigure[]{
             
            \includegraphics[scale=0.45]{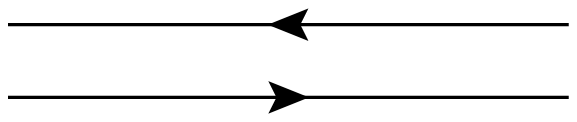}
              \put(-74,15){$j$}    
              \put(-74,-7){$i$}   
              \put(-3,15){$k$}    
              \put(-3,-7){$l$}   }   
            \hspace{1cm}    
            \\      
             \subfigure[]{ 
                      
              \includegraphics[scale=0.45]{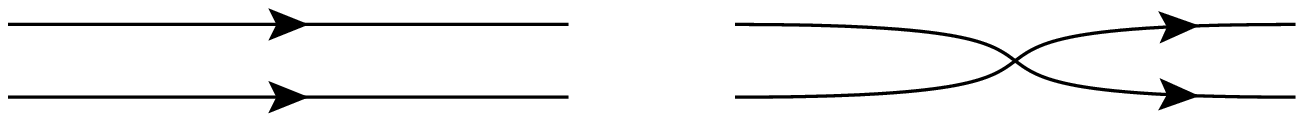}
              \put(-88,5){$-$}
              \put(-74,15){$j$}    
              \put(-74,-7){$i$}   
              \put(-3,15){$k$}    
              \put(-3,-7){$l$} 
              \put(-169,15){$j$}    
              \put(-169,-7){$i$}   
              \put(-98,15){$k$}    
              \put(-98,-7){$l$} 
              }
              \\      
             \subfigure[]{ 
                      
              \includegraphics[scale=0.45]{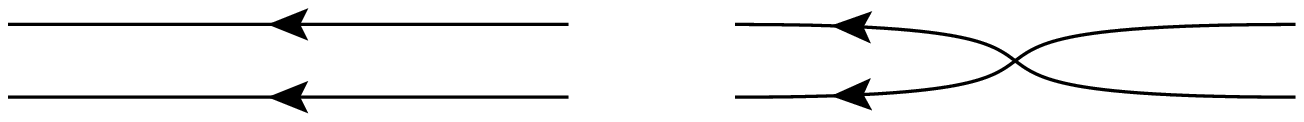}
              \put(-88,5){$-$}
              \put(-74,15){$j$}    
              \put(-74,-7){$i$}   
              \put(-3,15){$k$}    
              \put(-3,-7){$l$} 
              \put(-169,15){$j$}    
              \put(-169,-7){$i$}   
              \put(-98,15){$k$}    
              \put(-98,-7){$l$} 
              }
              \caption{Fermionic propagators in a double-line notation.}
              \label{propagators}
            \end{center}
            \end{figure}

		From Figure \ref{propagators} we see that the orientifold daughter has an extra ``non-planar'' term coming from the 
		antisymmetry of the color indices.
                                  
       We are only interested in double trace contributions. In particular,
       \begin{equation}
        \delta \mathcal{L} =
        a(\lambda)\mathcal{O}^{IJ}\mathcal{O}^{IJ}+b(\lambda)\mathcal{O}^2\, ,
        \end{equation}
        where
        \begin{equation}
        \mathcal{O}^{IJ}=\textrm{Tr}\left(X^IX^J-\frac{\delta^{IJ}}{6}X^KX^K\right)\,,
        \end{equation}
        \begin{equation}
        \mathcal{O}=\textrm{Tr}X^K X^K\,,
        \end{equation}
        and $a(\lambda)$, $b(\lambda)$ are functions of the 't Hooft coupling.
       
\begin{figure}[h]
             \begin{center}
             
             \subfigure[]{ 
                      
              \includegraphics[angle=90,scale=0.3]{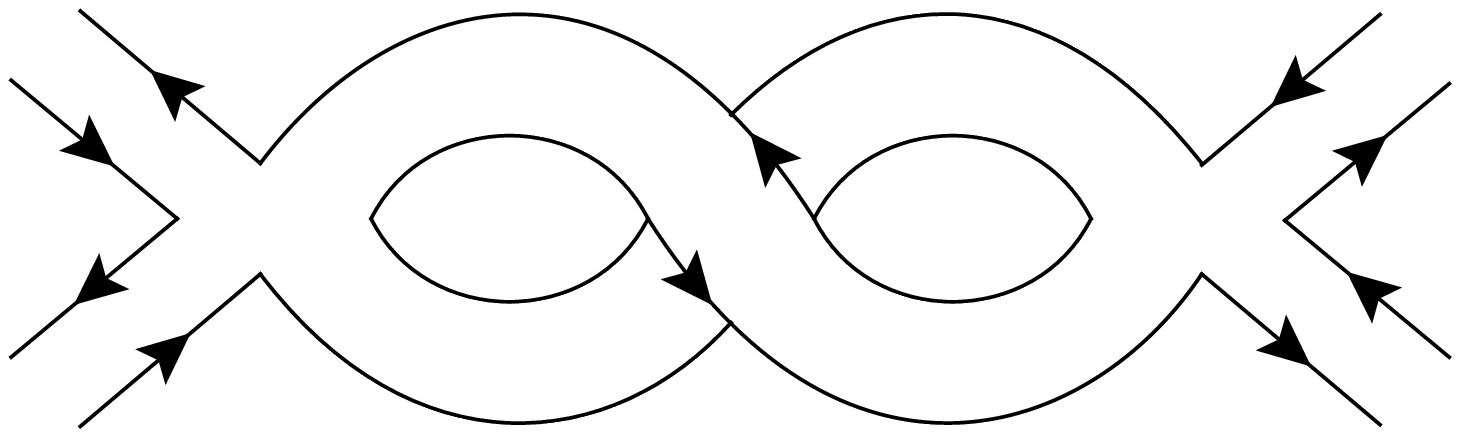}
              }
             \hspace{1cm}                     
             \subfigure[]{ 
                      
              \includegraphics[scale=0.45]{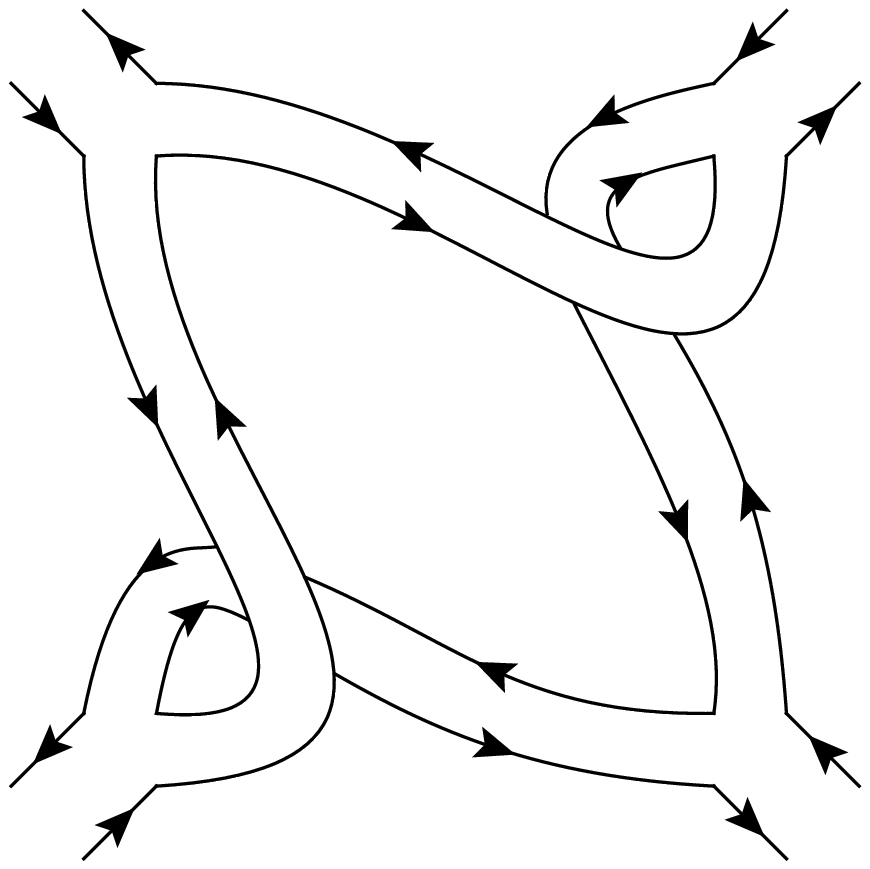}
            \put(-90,105){$\psi$}
            \put(-28,105){$\bar{\psi}$}
     		\put(-110,82){$\psi$}
     		\put(-9,82){$\bar{\psi}$}
     		\put(-110,22){$\bar{\psi}$}
     		\put(-9,22){$\psi$}
     		\put(-90,-1){$\bar{\psi}$}
            \put(-28,-1){$\psi$}
              }
              \hspace{1cm}          
             \subfigure[]{ 
                      
              \includegraphics[scale=0.45]{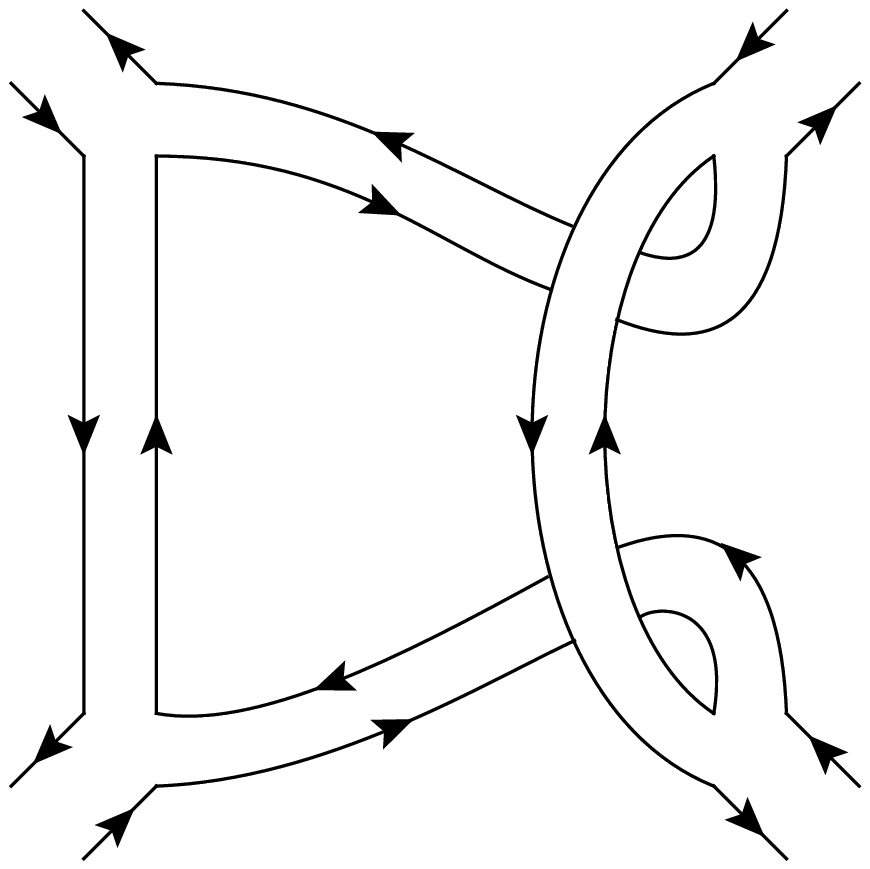}
            \put(-90,105){$\psi$}
            \put(-28,105){$\bar{\psi}$}
     		\put(-110,82){$\psi$}
     		\put(-9,82){$\bar{\psi}$}
     		\put(-110,22){$\bar{\psi}$}
     		\put(-9,22){$\psi$}
     		\put(-90,-1){$\bar{\psi}$}
            \put(-28,-1){$\psi$}
              }
              \caption{Figure (a) is a bosonic contribution to the double-trace potential. Figures (b) and (c) are fermionic contributions.}
              \label{sample}
            \end{center}
            \end{figure}

       These terms are easy to identify using the double-line notation, some
       sample contributions for $\Nm=4$ SYM are shown in Figure \ref{sample}. The bosonic diagrams are identical in both theories and 
       we know that the $\Nm=4$ SYM does not have double-trace running.
	    Then, we only need to concentrate in the fermions. There are two different sets of Wick contractions that give fermionic 
	    double-trace contributions in $\Nm=4$, Figures \ref{sample}(b) and \ref{sample}(c).
               
\begin{figure}[h]
             \begin{center}
            
             \subfigure[]{ 
                      
              \includegraphics[scale=0.45]{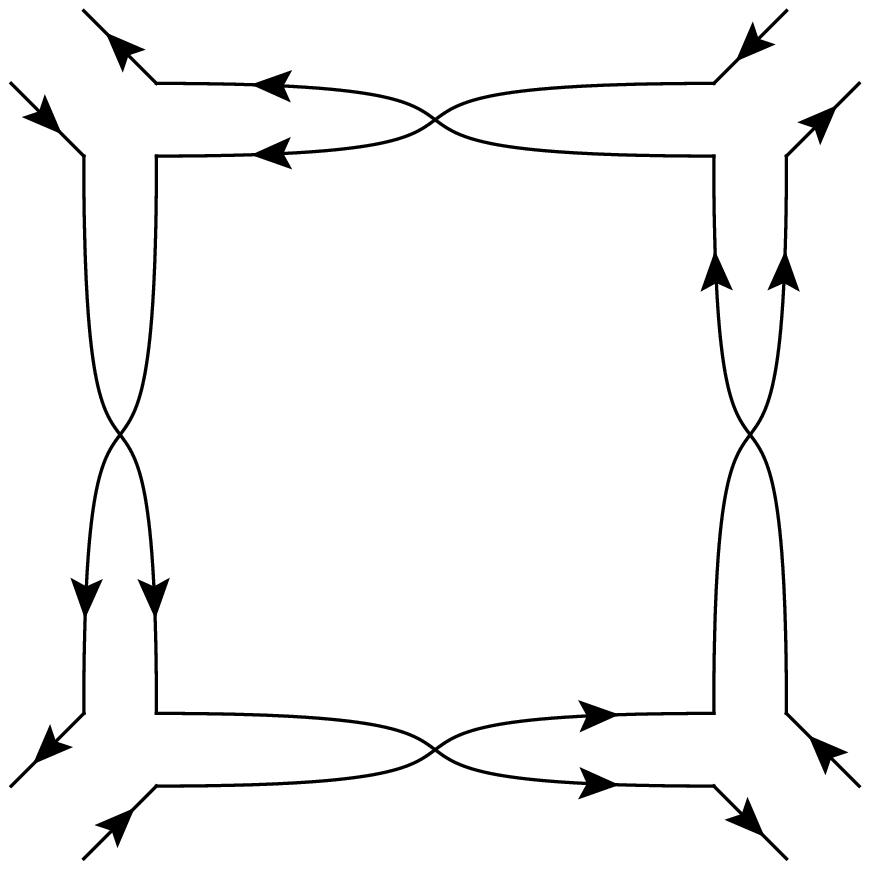}
              \put(-90,105){$\xi$}
            \put(-28,105){$\bar{\xi}$}
     		\put(-110,82){$\eta$}
     		\put(-9,82){$\bar{\eta}$}
     		\put(-110,22){$\bar{\eta}$}
     		\put(-9,22){$\eta$}
     		\put(-90,-1){$\bar{\xi}$}
            \put(-28,-1){$\xi$}
              }
                \hspace{1cm}     
             \subfigure[]{ 
                      
              \includegraphics[scale=0.45]{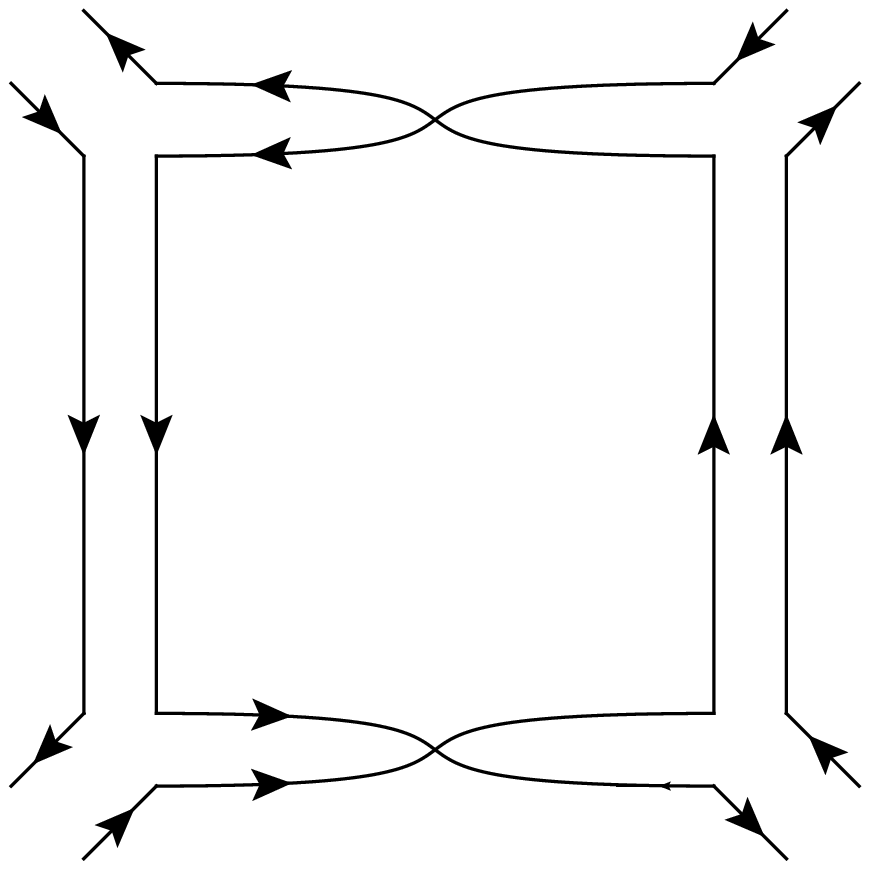}
              \put(-90,105){$\xi$}
            \put(-28,105){$\bar{\xi}$}
     		\put(-110,82){$\eta$}
     		\put(-9,82){$\bar{\eta}$}
     		\put(-110,22){$\bar{\eta}$}
     		\put(-9,22){$\eta$}
     		\put(-90,-1){$\bar{\xi}$}
            \put(-28,-1){$\xi$}
              }
              \caption{Fermionic one-loop contributions from the orientifold daughter.}
              \label{OrientiOneLoop}
            \end{center}
            \end{figure}

	    To prove cancellation of double-trace terms in the orientifold we need to find equivalent fermionic diagrams, these 
	    are shown in Figure \ref{OrientiOneLoop}. These two diagrams will give identical contributions as those of the parent theory 
	    and the cancellation of double-trace 
	    	 terms is then guaranteed. We see that the non-planarities in the $\Nm=4$ diagrams are here implemented by the non-
	    	 planar 
	    	 component of the fermion 
	    	 propagator. The extra minus sign in the non-planar part could have presented a problem, however, both diagrams have an 
	    	 even number of them. This simple analysis confirms that the orientifold theory has no double-trace 
	    	 contributions at one loop.
	    
	    \subsection{Two-loop example}
	    	    
\begin{figure}[h]
             \begin{center}
            
             \subfigure[]{
             
            \includegraphics[scale=0.35]{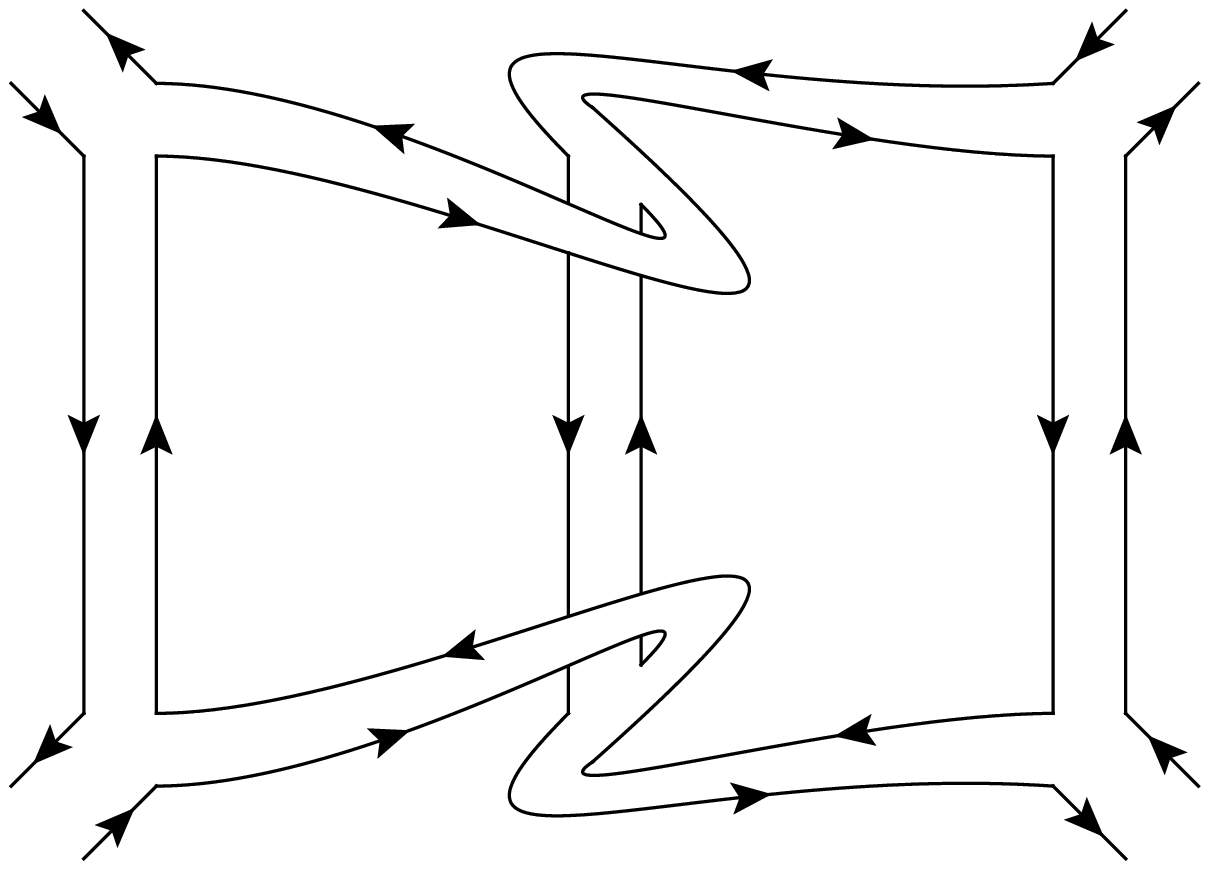}
                       }   
            \hspace{1cm}          
             \subfigure[]{ 
                      
              \includegraphics[scale=0.35]{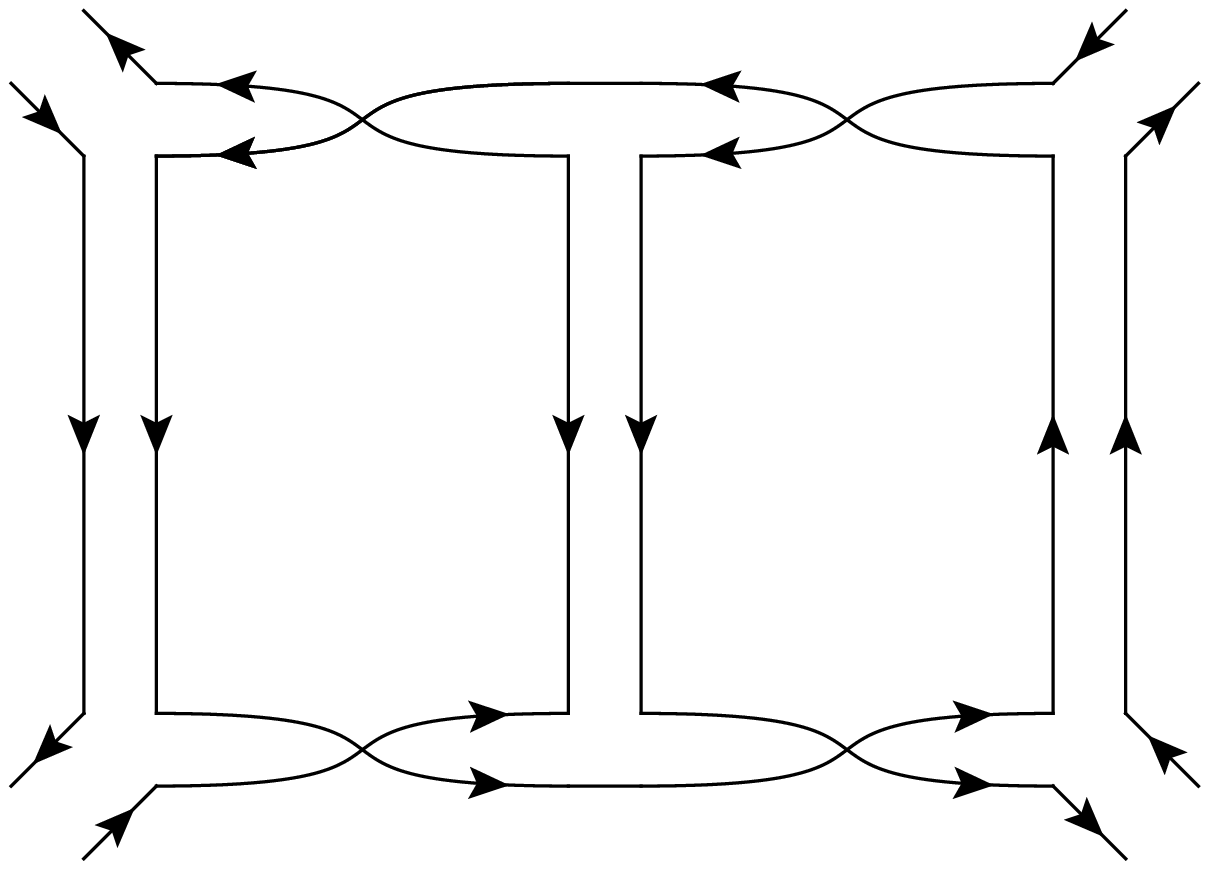}
                         }
             \caption{Figure (a) is a two-loop diagram in $\Nm=4$ SYM. Figure (b) is the orientifold counterpart.}
              \label{TwoLoop}
            \end{center}
            \end{figure}

	   At two-loops we should have one power of $N$ coming from a closed color loop, if not, the diagram is sub-leading. As before, 
	   we only 
	   need to 
	   concentrate on the diagrams with fermions. 	    
	    Because we will proceed with an all-loop analysis in the next section we only consider one two-loop example, this is shown in Figure \ref{TwoLoop}.
	    
        From the color flows is clear that we have a factor of $N$ coming from a closed color loop. Also, for the orientifold, we 
        have an even number of non-planar propagators and so 
	    there is no extra minus sign.

            
             
                      
             
                      
      	    
	    \subsection{All-loop}
	    
\begin{figure}[h]
             \begin{center}        
                  \subfigure[]{ 
                      
              \includegraphics[scale=0.5]{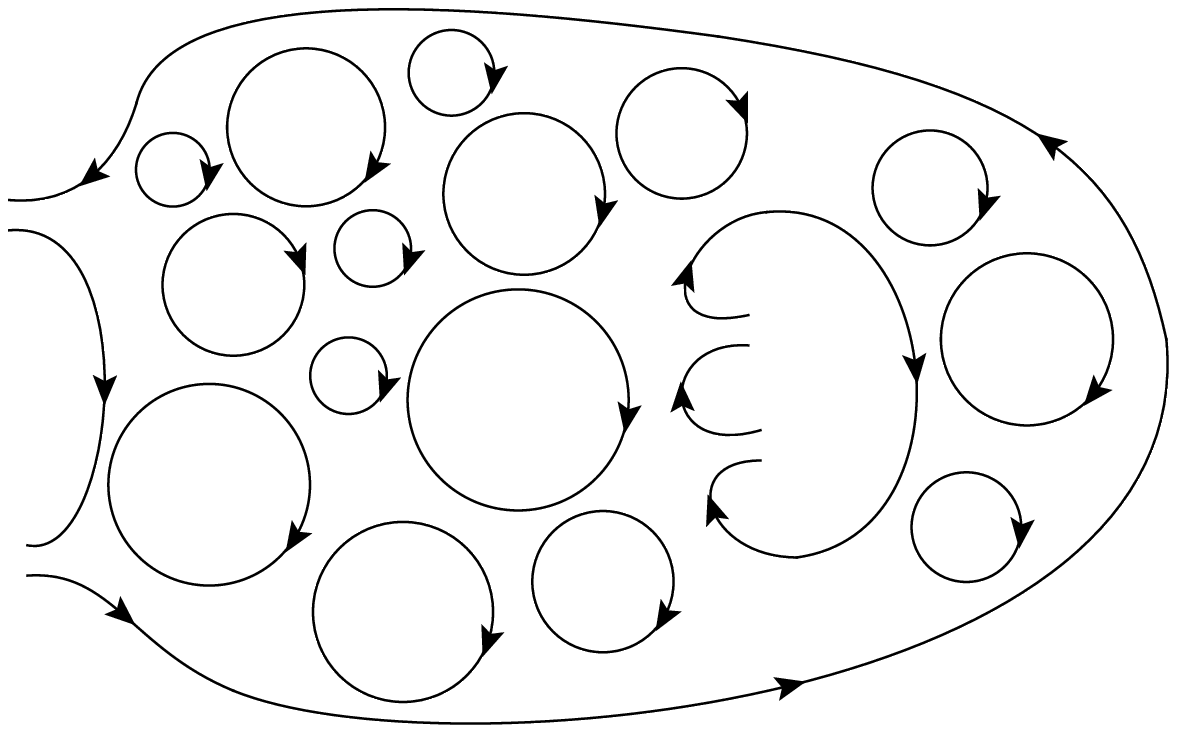}
              }
                \hspace{1cm}     
             \subfigure[]{ 
                      
              \includegraphics[scale=0.5]{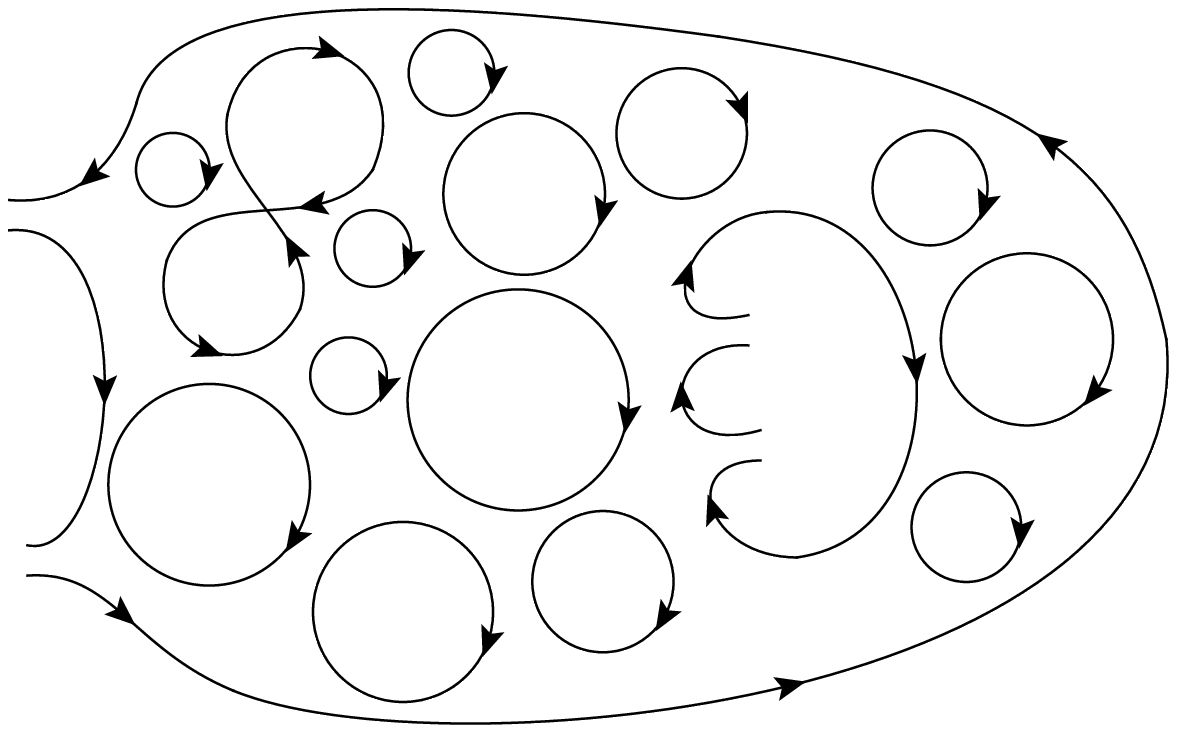}
              }   
             \caption{Figure (a) is a leading $\ell$-loop double-trace diagram, it has $N^{\ell-1}$ coming from closed color loops. 
             Figure (b) is sub-leading.}
              \label{LeadingDoubleTrace}
            \end{center}
            \end{figure}

        Here we will argue for the cancellation of double-trace terms at all loops.  At $\ell$-loops the leading single and 
        double-trace contributions are of the form,
          \be 
        \delta \Lm_{ST} \sim N \l^{\ell+1} \textrm{Tr}X^4 \hspace{1cm}\textrm{and} \hspace{1cm} \delta \Lm_{DT} \sim \l^{\ell+1}
        \textrm{Tr}X^2 \textrm{Tr}X^2\,.
        \ee        
        The key thing to notice is that a 
        leading double-trace diagram should have $N^{\ell-1}$ from closed color loops\footnote{In our normalization we have an 
        overall factor of $N$ multiplying the lagrangian, this implies that each propagator goes as $\sim \frac{1}{N}$.}. If we 
        compare this with the $N^\ell$ of a leading single-trace diagram it's clear that the double-trace contributions we 
        are interested in are almost planar. 
        The unique topology is shown in Figure \ref{LeadingDoubleTrace}(a). It's not hard to see that the double-trace diagrams 
        of the previous sections are of this form.
        
        To prove that the orientifold daughter does not suffer from double-trace running we will proceed as before. For each 
        double-trace contribution in $\Nm=4$ SYM we will show that there is an equivalent diagram in the orientifold and vice versa.         
               
\begin{figure}[h]
             \begin{center}

              \includegraphics[scale=0.6]{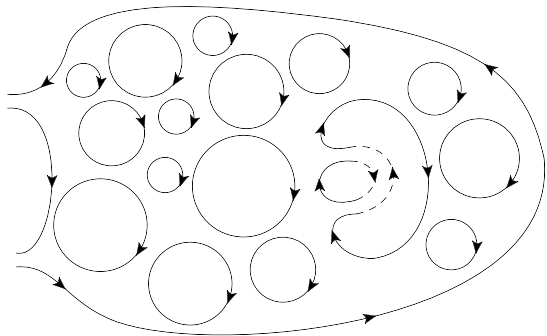}
               
             \caption{Leading $\ell$-loop double-trace diagram with an auxiliary bosonic propagator (dashed) contracting two external 
             legs (compare with Figure \ref{LeadingDoubleTrace}(a)).}
              \label{LeadingDoubleTraceContracted}
            \end{center}
            \end{figure}

        Consider an arbitrary leading double-trace diagram like 
        the one shown in Figure \ref{LeadingDoubleTrace}(a). Topologically we have an inner and an outer boundary where the 
        external bosonic legs sit and a number of color loops between them. Now, let's connect the external legs belonging to the 
        inner boundary using an ``auxiliary'' bosonic propagator 
        as shown in Figure \ref{LeadingDoubleTraceContracted}. 

        After this contraction we have a \textit{planar} diagram (the new topology consists of a single boundary with two external 
        legs) but we know that
        there is a one-to-one correspondence between the planar diagrams 
        of 
        these two theories \cite{Angelantonj:1999qg}. We also know that the bosonic fields are identical in both theories and the 
        only difference 
        between Figure \ref{LeadingDoubleTrace}(a) and Figure \ref{LeadingDoubleTraceContracted} is a bosonic contraction, this 
        implies that if 
        there is a one-to-one correspondence between the diagrams of the form depicted in Figure \ref{LeadingDoubleTraceContracted} 
        then, there is also a
         one-to-one correspondence between the class of diagrams depicted in Figure \ref{LeadingDoubleTrace}(a). Let's rephrase this 
         last statement: If we 
         have a 
        one-to-one 
        correspondence 
        between planar diagrams in $\Nm=4$ SYM and its orientifold daughter, then we also have a one-to-one correspondence between 
        the leading double-trace contributions to the scalar four-point function.
        This concludes our analysis and 
        confirms that both theories have identical leading double-trace contributions at all loops.

   \newpage

\section{Discussion}

In this note we have shown by an explicit diagrammatic analysis that the orientifold daughter of $\Nm=4$ SYM does not suffer from 
double-trace running. This is in agreement with the results of \cite{Dymarsky:2005nc} and \cite{Armoni:2007jt}, where a one-to-one correspondence was found between the breaking of conformal invariance and the presence of tachyons in the flat space string theory. Our calculation is yet another example that confirms this observation, namely that the flat space theory seems to \textit{know} about the stability of the field theory. 

In the full 0B string theory the calculation of \cite{Tseytlin:1999ii,Dymarsky:2005uh} implies that there is a string state becoming tachyonic in the dual AdS background for sufficiently small $\lambda$ (large curvature). The results of this paper confirm that the orientifold daughter is stable and hence the AdS dual contains no tachyon. It would be interesting to understand more directly how the tachyon is projected out. Sadly, the AdS background dual to the orientifold daughter is not known. In \cite{Armoni:2008kr} an outline was given of the main characteristics the dual theory should have. However, an explicit solution is yet to be found.

\begin{acknowledgments}

I am grateful to Leonardo Rastelli for suggesting this project and for guidance.
It is a pleasure to thank Elli Pomoni and especially Adi Armoni for useful discussions and comments.
This  work was supported in part by NSF grant PHY-0969919. Any opinions, findings, and conclusions or recommendations expressed in this material are those of the authors and do not necessarily reflect the views of the National Science Foundation.

\end{acknowledgments}

\newpage

\bibliographystyle{JHEP}
\bibliography{Completebbl}

\end{document}